\documentclass[10pt, conference,compsocconf]{IEEEtran}  

\usepackage[bottom]{footmisc}
\usepackage{ifpdf}
\usepackage{cite}
\ifCLASSINFOpdf        
   \usepackage[pdftex]{graphicx}
\else
   \usepackage[dvips]{graphicx}
\fi
\usepackage[cmex10]{amsmath}   
\usepackage{algorithmic}
\usepackage{array}
\usepackage{bm}
\usepackage{mdwmath}
\usepackage{mdwtab}
\usepackage{eqparbox}
\usepackage{url}
\usepackage{subfigure}
\usepackage{balance}
\usepackage{multirow}
\usepackage{array}       
\usepackage{multirow}  
\usepackage[numbers,square,comma,sort&compress]{natbib}   

\usepackage[draft,inline,nomargin]{fixme}   
\usepackage{color}
\usepackage{soul}          

\begin{document}

\title{Q-A: Towards the Solution of Usability-Security Tension in User Authentication}


\author{\IEEEauthorblockN{Mahdi Nasrullah Al-Ameen}
\IEEEauthorblockA{
The University of Texas at Arlington\\
mahdi.al-ameen@mavs.uta.edu}
\and
\IEEEauthorblockN{S M Taiabul Haque}
\IEEEauthorblockA{
The University of Texas at Arlington\\
eresh03@gmail.com}

\and
\IEEEauthorblockN{Matthew Wright}
\IEEEauthorblockA{
The University of Texas at Arlington\\
mwright@cse.uta.edu}
}

\maketitle
\balance
\begin{abstract}
Users often choose passwords that are easy to remember but also easy to
guess by attackers. Recent studies have revealed the vulnerability of
textual passwords to shoulder surfing and keystroke loggers. It remains
a critical challenge in password research to develop an authentication
scheme that addresses these security issues, in addition to offering
good memorability. Motivated by psychology research on humans' cognitive
strengths and weaknesses, we explore the potential of {\em cognitive
questions} as a way to address the major challenges in user
authentication. We design, implement, and evaluate Q-A, a novel
cognitive-question-based password system that requires a user to enter
the letter at a given position in her answer for each of six personal
questions (e.g. ``What is the name of your favorite childhood
teacher?"). In this scheme, the user does not need to memorize new,
artificial information as her authentication secret. Our scheme offers
$28$ bits of theoretical password space, which has been found sufficient
to prevent online brute-force attacks. Q-A is also robust against
shoulder surfing and keystroke loggers. 
We conducted a multi-session in-lab user study to evaluate the usability
of Q-A; $100$\% of users were able to remember their Q-A password over the
span of one week, although login times were high. We compared our scheme
with random six character passwords and found that login success rate in
Q-A was significantly higher. Based on our results, we suggest that Q-A
would be most appropriate in contexts that demand high security and
where logins occur infrequently (e.g., online bank accounts).

\end{abstract}

\begin{IEEEkeywords}   
Usable security; user authentication; cognitive question
\end{IEEEkeywords}



\section{Introduction} 

Compared with the tremendous advancement of digital devices over the
past few decades, passwords are a stubborn legacy technology that hangs
on for want of something better. 
Recall-based user-chosen textual password is the most widely used authentication scheme on the Web, but it is fraught with security problems because of password reuse~\cite{pwhabit} and predictable patterns~\cite{pwpattern1,pwpattern2} that make this scheme vulnerable to online guessing attacks~\cite{captcha,guessing08}. Password restriction policies are found to have limited impact on ensuring security and in some cases adversely affect password memorability~\cite{
furnell07,pwpattern1}. System-assigned random passwords provide higher security, but suffer from memorability problems~\cite{yan_sec}. Multiple variants~\cite{passphrase,text_recog,persuation,forget_thesis} have been proposed to improve memorability in this respect, yet none of these schemes has shown significant improvement for the examined usability
metrics.

Users have cognitive limitations that define our potential for
interaction with computers. Existing password systems fail to fully
address these limitations, nor do they leverage humans' cognitive
strengths.

\subsection{Motivation}  
The most prominent dilemma studied in the research on user
authentication is the delicate balance between memorability and
preventing successful guessing. Existing schemes rely on information
specifically memorized for the purpose of authentication, such as the
string of characters that make up a textual password. This means that
the information is less memorable for users than information that users
already know because it is meaningful for them. This in turn means that
users tend to choose passwords that are easy to guess.

Beyond guessing resistance, however, other security concerns have been
shown to be at least as important. Florencio et al. point
out~\cite{strongpw} that observation attacks, such as \textit{shoulder
  surfing} (external observation attack) and \textit{keystroke loggers}
(internal observation attack~\cite{design_space}), appear to be the most
prevalent attacks, in which strong passwords are just as susceptible to
being stolen by an attacker as weak ones. They find that none of the
password ``best practices'' offers any real protection against these
attacks. This observation is also supported by the study of Florencio
and Herley~\cite{lockout}, who examined the password policies of $75$
different websites and found that their password policies address only
guessing attacks and, in few cases, passwords reuse across sites.

The susceptibility to shoulder surfing and keystroke loggers is
generally high when users log in from public
computers~\cite{keylogger}. A two-week-long field study~\cite{pw_diary}
on real life Web usage found that half of the participants used public
computers to access online accounts.

Thus, one of the most challenging tasks in user authentication today is to develop a text-based authentication scheme that satisfies the following requirements: i) Provides protection against observation attacks, ii) Offers good memorability, and iii) Provides sufficient entropy to prevent online brute-force attacks.

\subsection{Contributions} 
In this paper, we explore the potential of {\em cognitive questions} in addressing these challenges by incorporating the scientific
understanding of long-term memory to advance the security and usability
properties that can be provided in an authentication system. 
In particular, we propose {\em Q-A}, a new authentication scheme in which a user registers by selecting and answering six cognitive
questions (e.g., ``Who was your favorite high school teacher?'') from a
set of twenty questions. During login, the user is shown one question at
a time and asked to enter the letter at a random position in her
answer. This process is repeated for all six questions, and correctly
entering all six letters is required to authenticate. The contributions
of Q-A are given below (see~\S\ref{design} for a detailed discussion).

\textbf{Usability.} A survey study~\cite{survey} on $25$ different
password schemes found that memorability can be supported by leveraging
pre-existing user-specific knowledge, rather than requiring users to
memorize new information that is artificially constructed or
random. This finding inspires our application of cognitive questions in
Q-A, which gains usability advantages over other text-based password
schemes~\cite{mnemonic,persuation,text_recog,forget_thesis} in the
following way:

\begin{itemize}
\item Authentication secrets are easier to remember in Q-A, since it
  asks users for already known information, while the other schemes
  query for specifically memorized information.
\item Psychology studies~\cite{grecog, kintsch, tulving73} reveal that
  it is difficult to remember information spontaneously without memory
  cues, where cognitive questions work as cues to retrieve the
  corresponding answers from a long-term
  memory~\cite{qa04,forget_thesis}.
\item Our scheme allows users to enter case-insensitive letters for
  authentication, while traditional textual passwords may contain
  numbers, upper-case letters, and special characters, which may require
  additional time and effort to enter, especially from portable devices
  (e.g., cellular phones).
\end{itemize}

In our study, users had a $100$\% memorability rate after one week, which
was significantly higher than that for randomly assigned six-character
passwords consisting of only lowercase letters. 
They reported high levels of satisfaction with the usability and
security of Q-A, despite a long average time to log in.

\textbf{Security.}  To the best of our knowledge, Q-A is the first
text-based authentication scheme that addresses both internal and
external observation attacks. The security features of our system are as
follows:

\begin{itemize}
\item Typically, cognitive questions are prone to \textit{guessing by acquaintances attacks} that exploit the knowledge about personal information of a user~\cite{qa_just09}. The analysis of Just and Aspinall~\cite{qa_just09} shows that \textit{three} questions are sufficient to guarantee reasonable security for authentication based on cognitive questions. Our study on Q-A satisfied these security requirements, where a set of questions were carefully selected considering both usability and security metrics~\cite{no_sec,qa_just09}, and each user was asked to answer \textit{six} questions from this set to measure the usability of Q-A for a system with high security requirements.
\item Q-A offers the \textit{variant response} feature, where users' entries vary across different login sessions to gain resilience against observation attacks.
\item Although each security question offers relatively low entropy compared with a moderately strong password~\cite{qa_just09}, selecting random letters from the answers to six different questions makes for a string of characters that is partially random and therefore hard to guess~\cite{yan_sec}. The theoretical password space ($28$ bits) offered by Q-A is sufficient to prevent online brute-force attacks~\cite{lockout}.
\end{itemize}

With Q-A, we explore the idea of using information that
is meaningful to users and already held in long-term memory for a
primary authentication mechanism. By developing and evaluating the Q-A
design, we show that such a mechanism can be built with reasonable
usability and security properties. We hope that this inspires
researchers to further investigate ways to leverage such information for
better schemes.

\section{Related Work}\label{related} 

In this section, we give an overview of the text-based authentication
mechanisms and the proposed techniques to gain resilience against
observation attacks.

\textbf{Traditional password.} Recall-based user-chosen textual password
is the most widely used authentication scheme on the web. However, it is
vulnerable to online guessing attacks~\cite{captcha,guessing08} that
exploit password reuse and predictable patterns, since users use the
same password on an average of six different accounts~\cite{pwhabit} and
follow predictable strategies (e.g., using dictionary words or names)
while creating passwords~\cite{pwpattern1,pwpattern2}.

To motivate users to create stronger passwords, different password
restriction policies~\cite{pwpattern1,pwmeter13} have been deployed,
such as increasing minimum length of passwords, using combination of
different types of characters, asking users to change passwords at
regular intervals, and using password strength meters. However, in
separate studies, 
Proctor et al.~\cite{proctor_sec} and Shay et al.~\cite{pwpattern1} reported that
password restriction policies do not necessarily lead to more secure
passwords. Rather, in some cases, they adversely affect the
memorability.

\textbf{Mnemonic Password.} Kuo et al.~\cite{mnemonic} studied the
guessing-resistance of user-selected mnemonic phrase based passwords, in
which the user chooses a memorable phrase and uses a character (often
the first letter) to represent each word in the
phrase. Results~\cite{mnemonic} show that user-selected mnemonic
passwords are slightly more resistant to brute-force attacks than
traditional text-based passwords, . Mnemonic passwords are found more
predictable when users choose common phrases to create their
passwords. A properly chosen dictionary may increase the success rate in guessing mnemonic passwords~\cite{mnemonic}.

\textbf{System-assigned password.} Randomly assigned textual passwords
provide better resilience against online guessing attacks in comparison
to user-chosen text-based passwords~\cite{yan_sec}. Wright et
al.~\cite{text_recog} compared the usability of three different
system-assigned textual password schemes: i) Recognition-based words,
ii) Recall-based words, and iii) Recall-based random letters. In the
recognition-based scheme, the user has to recognize a set of words, each
word being displayed on screen amongst a set of distracter words. In the
second scheme, users have to remember a list of four whole words, which
serves as one password. In the last scheme, users are assigned random
passwords of six lower-case letters. The password space for all the
conditions are kept same ($28$ bits). Results for memorability show that
word recall performs the worst and no significant difference is found
between recognition and letter recall. Furthermore, the time required to
login is significantly different for all pairings, and notably worst in
the recognition condition.

\textbf{System-assigned passphrase.} Shay et al.~\cite{passphrase}
investigated the usability of system-assigned passphrases
(space-delimited sets of natural language words), which did not show
significant improvement over the system-assigned textual passwords of
similar entropy. The study~\cite{passphrase} indicates that a majority of participants in each condition wrote down their passwords/passphrases, and around half of the participants who did
not write them down failed to recall their authentication secrets after
two days.

\textbf{PTP.} Forget et al.~\cite{persuation,forget_thesis} proposed the
Persuasive Text Passwords (PTP) scheme, in which, after a password is
created by the user, PTP improves its security by placing
randomly-chosen characters at random positions into the password. Users
may shuffle to re-position the random characters until they find a
suitable combination to memorize. PTP is resilient against attacks
exploiting password reuse and predictable patterns. However, knowing that a system uses PTP and knowing how PTP
works would allow attackers to refine their cracking strategies. The
memorability for PTP is found to be just $25$\% when two random
characters are inserted at random positions~\cite{forget_thesis}.

\textbf{Cognitive question.} The primary investigation of Furnell
et al.~\cite{qa04} reported that $70$\% users preferred to have
cognitive questions for authentication. However, most of the questions
used in their study (e.g., ``What is the name of your favorite
relation?", ``What is your favorite shape?") offer very limited entropy
and thus can be easily guessed by attackers. Also, their approach is
vulnerable to shoulder surfing and keystroke loggers. Later
studies~\cite{no_sec,qa_just09,qa_fb} performed comprehensive analysis
on cognitive questions to figure out the usability and security of
different types of questions. We have carefully considered their findings to
design Q-A. A detailed discussion follows in the next section.

\textbf{Resilience to observation attacks.} Textual passwords are vulnerable to {\em external observation attacks}, such as shoulder surfing~\cite{shoulder06}. De Luca et
al.~\cite{shoulder_chi13} examined the impact of fake cursors in
providing resilience against shoulder surfing when passwords are entered
through an on-screen keyboard. However, results show that the users act
predictably to identify their active cursor (e.g., moving the mouse
cursor to the border of the interaction area or moving the mouse in
small circles), which may make it easy for the shoulder surfer to find
the real cursor and subsequently the authentication secret from
on-screen keyboard entries.

Gaining the user's credentials through malware (e.g., keystroke loggers,
mouse loggers) is called an {\em internal observation
  attack}~\cite{design_space}. Text-based passwords are prone to
keystroke loggers~\cite{keylogger}. A few tricks have been proposed to
hide passwords from keystroke loggers, such as typing fake characters in
multiple text-boxes at the time of entering the
password~\cite{keylogger}. These tricks, however, require users to be
conscious about keystroke loggers and take proactive measures, which is
difficult~\cite{pwhabit,pwpattern2}.

\textbf{Conclusion and open problems.} As we see from the above discussion, textual password schemes that
provide better memorability are vulnerable to guessing attacks (e.g.,
traditional and mnemonic passwords). On the other hand, text-based passwords with higher resilience against guessing attacks suffer from memorability problems (e.g., system assigned passwords and passphrases,
PTP). Moreover, textual passwords are vulnerable to observation attacks. 

{\em Thus, despite a large body of research, it remains a critical challenge in password research to build an authentication system that offers both high memorability and guessing resilience, with robustness against observation attacks.}

\section{The Q-A Design and Scientific Motivations}\label{design}

\begin{figure}[t]
\centering
\includegraphics[width=70mm, height=25mm]{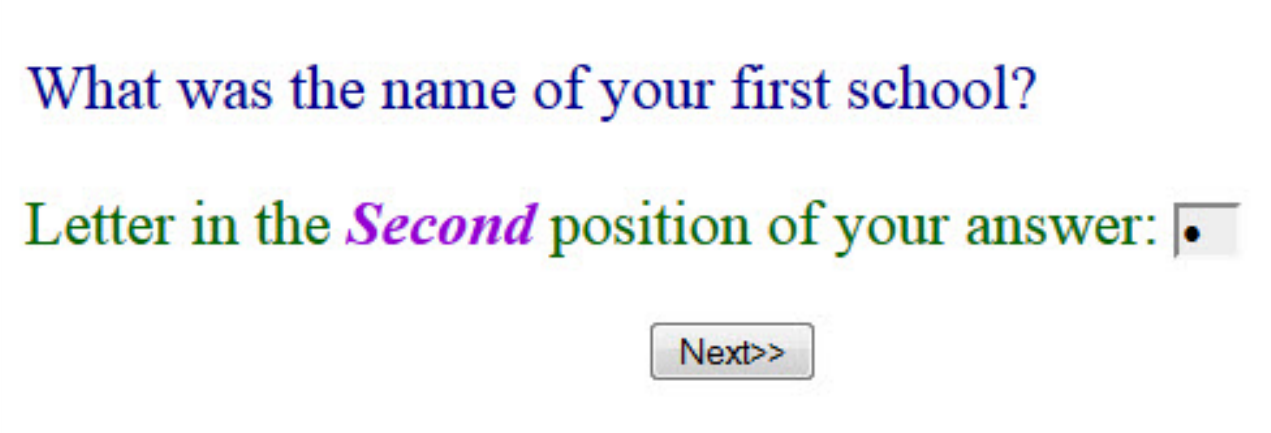}
\vskip -15 pt
\caption{A screenshot of a demo of the Q-A.}
\label{fig:ques}
\end{figure}

In this section, we first explain the basic design of Q-A. We then describe our approach to address the usability and security concerns with cognitive questions, followed by a detailed discussion on the usability and security features offered by Q-A.

Q-A is based on cognitive questions (e.g., ``What is the name of your favorite childhood teacher?''), which inherently leverage existing long-term memories, information that is known to users based on their
life experience. Q-A invokes the answers to cognitive questions in a novel way. At the time of registration, the user is shown a set of $20$ carefully selected questions, of which she must select six
questions to answer. These answers, in total, constitute her authentication secret. We choose the $20$ questions carefully to ensure that each one asks for alphabetical answers with a high amount of entropy, such as the name of a person or a location.

During login, a user is presented with all six selected questions, one at a time. She does not have to enter the whole answers. Instead, for each question, the user is asked to enter a single letter from a
particular position in her answer. Every time a user logs in, this position is randomly chosen by the system individually for each question. So, in a login session, a user may be asked to enter the letter in the \textit{second position} for the first question, in the \textit{fourth position} for the second question, and so on. These positions will vary in the subsequent login sessions. For example, if a user's answer is ``Anderson", and at a given login session she is asked to enter the second letter of this answer, she has to enter `n'. In this way, a user has to correctly enter a letter for all six questions for a successful login. 

\subsection{Usability and Security of Cognitive Questions}
In this section, we describe our approach in Q-A to address the usability and security concerns with cognitive questions.

\subsubsection{Usability} 

Just and Aspinall defined three metrics to measure the usability of a
cognitive question: i) Applicability, ii) Memorability, and iii)
Repeatability~\cite{qa_just09}. This provides a useful guide to
selecting appropriate questions for Q-A. We carefully considered these
metrics while selecting questions for our study.

\textbf{Applicability.} Not every user can effectively answer every
cognitive question. Users in our system choose any six questions from a
set of twenty questions that they find most applicable to them. Offering
a greater number of questions would further increase the applicability
of the system.

\textbf{Memorability.} Research shows that a user can easily recall the
answers of cognitive questions that are related to her long-term
memory~\cite{no_sec,qa_just09}. Selecting such questions ensures that
the user does not need to devote much cognitive effort in learning the
Q-A passwords, as they are simply answers that she already knows.

A study by Schechter et al. reveals that the participants who used weak
placeholder answers during the study failed to recall them
later~\cite{no_sec}. So, with a few basic restrictions
in place to guard against poor answers (see~\S\ref{sec_adv}), it should
typically be easier for a user to remember the real answer to a question
than a weak placeholder answer (`Aab').

The study of Furnell et al.~\cite{qa04} shows that confusion between
capital and lowercase letters is a prominent reason for making mistakes
when answering a cognitive question. We address this by ignoring case in
our scheme.

\textbf{Repeatability.} Just and Aspinall indicate that repeatability
can be improved by providing users with the fixed format of answers to
the questions that ask about dates or locations (e.g., different formats
for date: `Feb-05, 1992', `02-05-1992',
`02/05/1992')~\cite{qa_just09}. 
To provide higher entropy, we recommend to avoid questions that ask for
numerical answers, and this includes questions about specific dates. For
location-related questions, instead of imposing any specific format on
users, we prefer to be more specific with the questions to ensure
repeatability. For example: instead of asking ``Where did your father
and mother meet?'', we would ask ``In what city or town did your father
and mother meet?'' We recommend providing a fixed format for the answer
to a question in which being specific with the question does not resolve
the repeatability issues.

\subsubsection{Security} 
The most important security concerns with cognitive questions include:
i) User-created questions, ii) Guessing by acquaintances attacks, and
iii) Common answers. We address these security issues in the following
way.

\textbf{User-created questions.} If users are allowed to freely create their own questions, many users will not choose sufficiently secure questions~\cite{qa_just09}. On the other hand, it may have usability concerns for many users if the questions are strictly assigned by the system. Q-A balances these trade-offs by asking users to select any six questions from a larger set.

\textbf{Guessing by acquaintances.} Typically, cognitive questions are prone to targeted guessing attacks, in which attackers exploit the knowledge about personal information of a user~\cite{qa_just09,no_sec}. One's mother's maiden name and Social Security Number (SSN) in the US are particularly well-known examples of such questions. It is possible to select suitable questions, but still the amount of entropy in a single answer is typically lower than for a password~\cite{no_sec,qa_just09,qa_fb}.

Just and Aspinall~\cite{qa_just09} show that {\em three} questions are sufficient to guarantee reasonable security for cognitive-question-based authentication. In our study, we asked users to answer {\em six} questions to measure the usability of Q-A for a system with high security
requirements.

\textbf{Common answers.} The answers to some cognitive questions are generally common among users. For example, \textit{blue} or \textit{pink} may be common answers to the question ``What's your favorite color?". The prior studies~\cite{no_sec,qa_just09} have found that carefully selected questions can make common answers less of an issue for most users. The questions for Q-A were selected carefully based on the prior usability and security analysis on cognitive questions~\cite{no_sec,qa_just09,qa04,qa_fb}.

\subsection{Usability Features and Memory Retrieval in Q-A}\label{use_adv}

In this section, we state the usability advantages of Q-A from the perspective of cognitive psychology. Q-A offers two important usability advantages over other text-based authentication systems~\cite{mnemonic,persuation,text_recog,forget_thesis}:

\begin{itemize}
\item \textbf{Known Information:} Q-A queries for already known
  information, while the other schemes query for specifically memorized
  information.
\item \textbf{Memory Cues:} Q-A provides users with the questions that
  work as cues to retrieve the corresponding answers from user's memory,
  so as to log in successfully.
\end{itemize}

\textbf{Known Information.} From the viewpoint of cognitive
psychology~\cite{episodic72,encoding}, Q-A is closely tied with the
concept of \textit{episodic memory}~\cite{episodic72,encoding}, which
refers to the autobiographical event that the user can accurately recall,
since she was part of it. Episodic memory incorporates the time and place of a personally meaningful event with the associated feeling and contextual information. For example,
recalling the celebration of new millennium's eve involves figuratively
traveling back in time to precisely remember the place and the people
associated with that event. Thus, cognitive-question-based
authentication systems like Q-A aid password memorability, since a user
does not need to memorize the answers specifically for authentication to
an online account.

\textbf{Memory Cues.} Psychology research has shown that it is difficult
to remember information spontaneously without memory cues~\cite{grecog,
  kintsch, tulving73} . This suggests that authentication schemes should
provide users with cues to aid memory retrieval. Ellis and Kvavilashvili
state that memory cues support \textit{prospective
  memory}~\cite{memory09}, which is the ability to generate, retain, and
later recall information in the appropriate
context. \textit{Encoding specificity theory}~\cite{encoding} postulates that the most effective cues are those that are present at the time of remembering. If semantic information about a
word is processed at the time of learning, then that information can
successfully be used to cue memory. Thus, the word `Millennium Eve' can
only be used to cue memory of the word `New York' if the subject encodes
the semantic information linking the two objects at the time of
encoding.

\textit{Generate-recognize theory}~\cite{grecog} and
\textit{Associative-strength theory}~\cite{ellis89} also focus on the
effectiveness of cues in aiding memorability~\cite{kintsch,
  tulving73}. Generate-recognize theory~\cite{grecog}
speculates that retrieval is a two-step process, where in the generate
phase, a list of candidate words is formed by searching the long-term
memory. Then in the recognize phase, the list of words is evaluated to
see if they can be recognized as the sought out memory. According to
this theory, a cue can help not only in generating a relevant candidate
list, but also in recognizing the appropriate word from that
list. Associative-strength theory~\cite{ellis89} states that a cue
becomes effective if it has previously occurred with the remembered
event in the past. The theory assumes that memory is structured as a
network that connects all items in memory, and items in memory with
stronger ties between them make better cues.

\subsection{Security Features in Q-A}\label{sec_adv}

In this section, we describe the security features offered by Q-A.

\subsubsection{Guessing resistance} The theoretical password space ($28$ bits) offered by Q-A is sufficient to prevent online brute-force attacks~\cite{lockout}.

\textbf{Theoretical password space.} The questions in our scheme ask users for alphabetical answers (e.g., ``what was the name of your first teacher?"), where the answers are case-insensitive. During login, the user is asked to enter the character at a given position in her answer. In this respect, the size of the domain for an alphabetical entry ($26$) is larger than a numerical entry ($10$). Thus, alphabetical answers provide higher resilience against online brute-force attacks~\cite{guessing07,guessing08}, in which an attacker tries all elements within a search space to obtain the password. The user has to enter a letter for each of six questions. The space for this condition will therefore be $log_{2} {(26)}^6\approx 28$ bits. Florencio and Herley's study on security policies~\cite{lockout} suggests that $20$ bits of theoretical password space suffices for an online environment with lockout rules. 

\textbf{Effective password space.} Both in Q-A and in the study of password policies, only theoretical entropy is considered, though letters from random positions in the answers to cognitive questions may also have more effective entropy than user-selected passwords prone to dictionary attacks. In Q-A, the effective password space may vary for different questions, which requires to consider variation in answers in addition to general letter frequency. For example, the letter frequencies of names of people and locations will vary significantly by country and region. It would be an interesting area for future work to analyze the effective entropy of Q-A considering the letter frequency found in answers to different questions and the variation in answers in different languages, ethnicities, and countries.

\textbf{Guard against poor answers.}  We have deployed some basic restrictions in our system to guard against poor answers (e.g. minimum three characters in an answer, no repeat answers between questions, at
least two different letters in an answer), which might motivate users to give their real answers that, with the right questions, should have a large space of both possible and probable answers.

\subsubsection{Observation attacks} 

Having authentication information vary across login sessions is known as {\em variant response}~\cite{survey}. During login, the variant response feature makes our system more resilient against observation attacks like shoulder surfing, as compared to other text-based password schemes where the same set of characters is entered as password in every login session.

\textbf{Shoulder surfing.} When the user enters her credentials, either at registration or login, the answers are shown as asterisks or dots (as with regular password entry) to minimize the risk of shoulder surfing. During login, a user enters the letter at a given position in her answer, and to learn that letter and its position in the answer, the shoulder surfer needs to observe both the monitor and keyboard at a time, which has been found to be difficult in practice~\cite{shoulder06}. The requirements to observe the letters for all six questions further increase the hurdles for an attacker.

Even if a shoulder surfer can learn the letter and its position, he is likely to be asked to enter a letter of different position when he tries to log in as the user. Only a good guess of the entire answer for all six questions gives the attacker a reasonable chance of logging in.

The shoulder surfer may attempt to gain the user's credentials when she enters the entire answer to a question at the time of registration. Although answers are shown as asterisks or dots (as with regular password entry) to reduce the risk of shoulder surfing, we recommend that the users register in a secure environment (e.g., avoiding public terminals) to ensure maximum security.

\textbf{Keystroke and mouse loggers.} Gaining a user's authentication credentials through malware, such as keystroke loggers and mouse loggers, is called an {internal observation attack}~\cite{design_space}. A system provides resilience against keystroke/mouse loggers when the keyboard/mouse entries for authentication vary across subsequent login sessions~\cite{survey}. Thus, the variant response feature in Q-A offers higher resilience against keystroke loggers as compared to a password system where the same set of letters is entered (using keyboard) during each login session. Our system is clearly resilient to mouse loggers, as we don't use mouse input.

The security of Q-A can be further improved by asking users to answer more than six questions during registration and then drawing six at random for each authentication attempt. In this case, when an attacker tries to log in as a user, he may get different questions than the ones he gains through observation attacks. We will explore the usability of this approach in future work.

\subsubsection{Social engineering attacks}  

Social engineering refers to the psychological manipulation of people so that they divulge confidential
information~\cite{survey}. Phishing~\cite{survey} is a common form of social engineering attack.

\textbf{Phishing.} In a phishing attack, users are redirected to fraudulent websites to enter their credentials. In this case, the phishing victim will very likely get different questions from the ones she normally uses to log in, since a phisher would not typically have access to the user's login sessions. This means that not only will the user enter information that is useless to the attacker, she may realize that something is wrong and end the session. 

\textbf{Fake calls.} Users may be tricked to reveal credentials by any means, e.g., phone calls from a fake help desk or credit company~\cite{survey}. Q-A does not provide direct resilience to social engineering when the users disclose their authentication secrets to the attacker. However, a site with high security requirements can use unique questions to avoid question and answer reuse and overlap between sites so that attackers cannot gain access to other accounts by acquiring password for a single account. We propose to explore this issue further in future work.

\section{Study Design}\label{study}

Two prior text-based authentication systems~\cite{passphrase,text_recog} aim to provide a usable solution to the security issues with passwords. Both papers compared their scheme with system-assigned random textual passwords (\textit{control passwords}), where the results did not show significant improvement over control passwords in terms of memorability. We compare Q-A with control passwords and keep the theoretical password space the same ($28$ bits) for both conditions to examine whether Q-A offers better memorability in this respect.

In this study, we used a within-subjects design consisting of two experimental conditions (e.g., Q-A and control passwords). Using a within-subjects design controls for individual differences, and permitted the use of statistically stronger hypothesis tests. The experiments performed as part of this research received approval from our university's Institutional Review Board (IRB) for human subject research.

\subsection{Participants, Apparatus and Environment} 

For this experiment, we recruited $22$ university students ($6$ women, $16$ men) from diverse backgrounds, including majors from Biology, Engineering, Interdisciplinary Study, etc. The mean age of the participants was $27$. They make regular use of the Internet and websites that require authentication. Each participant was compensated with a \$10 gift card for participating in this study.

To administer this experiment, we created two realistic and distinct websites outfitted with the password scheme according to our two conditions: Q-A and control passwords. For both conditions, we used banking sites, but designed the interfaces with a different look and feel. Upon successful login at each bank site, the participants were forwarded to a dummy account overview page, to give them the feeling of
using online banking services. 

We carefully reviewed the prior usability and security studies on cognitive questions~\cite{no_sec,qa_just09,qa04,qa_fb} to select the $20$ questions for this study. The lab studies were conducted with one participant at a time to allow the researchers to observe the user's interaction with the system.

\subsection{Procedure} 

We conducted the experiment in two sessions, each lasting around $30$ minutes. The second session took place one week after the first one to test memorization of the password. Note that the one-week delay reflects a common interval used in authentication studies (e.g.,~\cite{face_age,text_recog,bdas}). A field study~\cite{pw_diary} on real-life web usage found that one week is larger than the maximum average interval for a user between her subsequent logins to any of her important accounts~\cite{pw_diary}. Thus, we used this interval to examine the usability of our scheme.

\subsubsection{Session 1} 

Before starting the experiment, we provided a consent form for the participants
to read and sign, if they agreed. To compensate for the novelty effect, we asked the participants to perform one practice trial for authentication with Q-A. We did not collect data for this
practice trial. The subsequent steps of the experiment are as follows:

\textbf{Sign-up.} During sign-up (we alternatively use the term {\em registration}) with Q-A, participants were shown a set of twenty questions, and they selected any six questions to answer, which constituted their password. Participants were also asked to sign-up with another site, where they were assigned a random six character password (control password).

To control for order effects, we employed counter balancing, so that all the participants would not see the schemes in the same order. Half of the participants used Q-A first, and the other half used the control password first.

\textbf{Distraction.} After sign-up, the participants were asked to count down in threes from a randomly chosen four-digit number for $45$ seconds. This type of distraction flushes the textual working memory~\cite{short_mem} and simulates a longer passage of time by focusing their attention on a separate, cognitively-difficult task. Participants were then given questionnaire that gathered demographic information.
 
\textbf{Login (Recall-1).} The participants were asked to log into each of those same sites, to demonstrate that the passwords had been memorized. Participants who were unable to reproduce their passwords during login, were shown the passwords. 

\subsubsection{Session 2} 
The second appointment took place one week after the first one. The participants were asked to log into each of the two sites (\textit{Recall-2}). After they had finished, an anonymous paper-based survey was conducted to get their opinion on the overall experiences of using the authentication schemes.  Participants were then compensated and thanked for their time.

\section{Results}

In this section, we discuss the results of the user study described
in~\S\ref{study}. We label the login performance of participants in
session 1 and session 2 as \textit{Recall-1} and \textit{Recall-2},
respectively. Here, we tested the following hypotheses:

\subsection{Hypothesis 1}

\vspace{0.03cm}
\noindent $H1_a$: \textit{The login success rate for Q-A and control passwords would not significantly differ in Recall-1.}
\vspace{0.03cm}

\noindent $H1_b$: \textit{The login success rate for Q-A would be significantly higher than that of control passwords in Recall-2.}
\vspace{0.05cm}

In Q-A, users don't have to memorize any new authentication secrets, as
their password instead comes from cognitive questions related to their
real life. However, in the control condition, the user is required to
memorize a random string of characters as her password. In Recall-1,
users were asked to enter their control passwords within a short period
of learning it. So, we hypothesized that the login success rate for Q-A
and control passwords would not significantly differ in Recall-1, but
that Q-A would perform significantly better than control passwords in
Recall-2, in terms of login success rate.

We observed a $100$\% login success rate for Q-A in both Recall-1 and
Recall-2. In the control condition, login success rate was $91$\% in
Recall-1 and $77$\% in Recall-2. Whether or not a participant
successfully authenticated is a binary measure, so we use McNemar's
tests when we analyze the login success rate for our within-subjects experiment. 
Our analysis shows that login success rate did not differ significantly between Q-A and control
conditions in Recall-1, $\mathcal{X}^{2}(1, N = 22) = 0.5$, $p =
0.24$. In Recall-2, however, the login success rate for Q-A was
significantly higher than the control passwords, $\mathcal{X}^{2}(1, N =
22) = 3.2$, $p<.05$. $H1_a$ and $H1_b$ are supported by these results.

\begin{figure}[t]
\centering
\includegraphics[width=85mm]{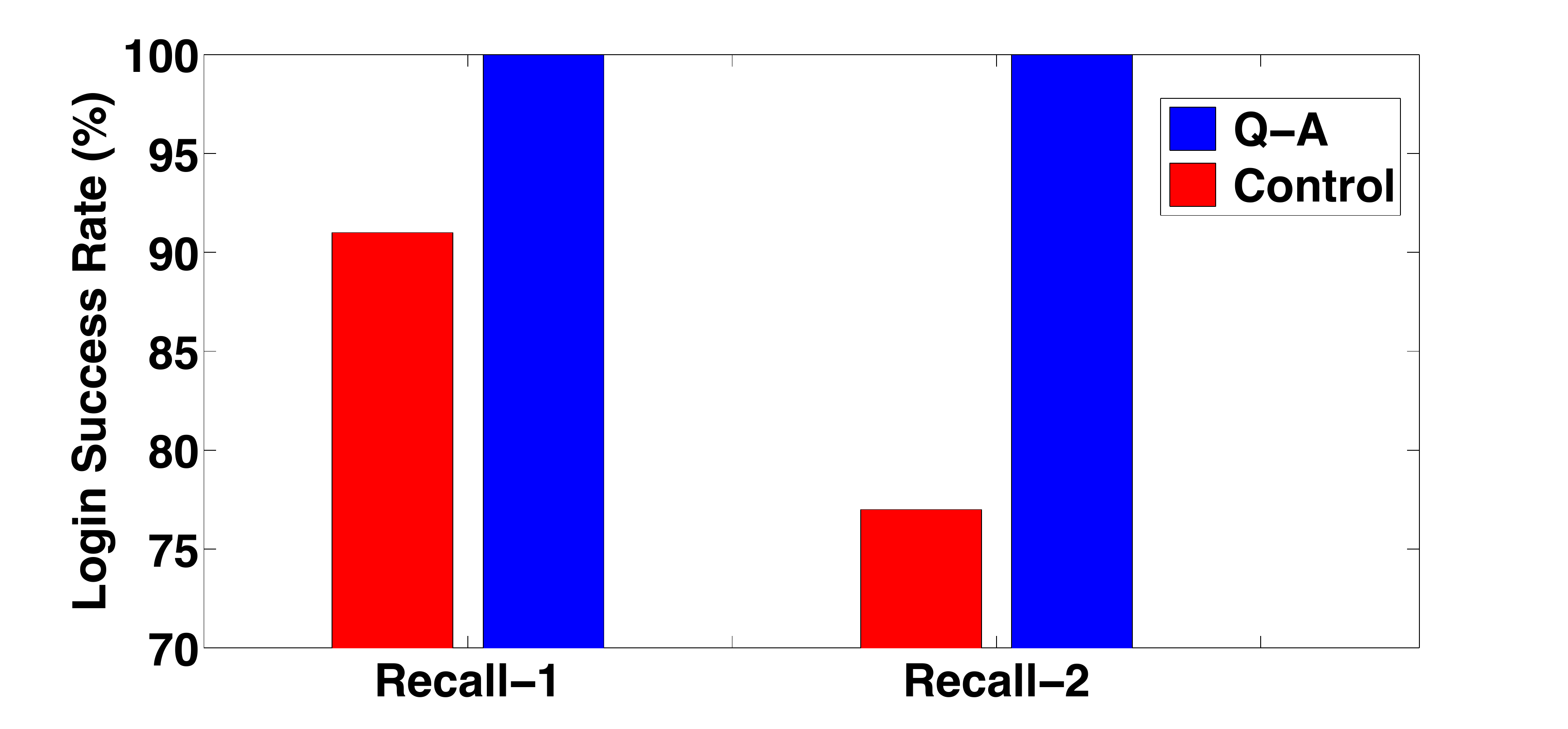}
\vskip -15 pt
\caption{Login success rate}
\label{fig:success}
\end{figure}

\subsection{Hypothesis 2}
\vspace{0.05cm}

\noindent $H2$: \textit{There would be a significant difference in login time between Q-A and control passwords in both Recall-1 and Recall-2.}
\vspace{0.05cm}

Since Q-A requires entering the letter of a random position for as many
as six answers, we hypothesized that for both Recall-1 and Recall-2, the
login time for the control passwords would be significantly less than
that of Q-A. Table~\ref{tab:time} summarizes the results for login time,
which shows that the mean login time in Q-A was $53.9$ seconds in Recall-1 (median: $51$ seconds) and $56.9$ seconds in Recall-2 (median: $53$ seconds). The mean login time ($38.1$ seconds in Recall-1, $43.7$ seconds in Recall-2) in the control condition seems to be affected by the outliers, where the median login time was $11.5$ seconds in Recall-1 and $13$ seconds in Recall-2.

We did not get matched pair of subjects while comparing the time for successful logins in our study conditions, since a number of participants who succeeded to log in using Q-A, failed in control condition. So, instead of Wilcoxon signed-rank test, we used a Wilcoxon-Mann-Whitney test to evaluate the difference in login time between Q-A and control passwords. Wilcoxon tests are similar to t-tests, but make no assumptions about the distributions of the compared samples, which is appropriate to the count data in these conditions. In Recall-1, we found a significant difference in login time ($W=72$, $p<0.01$). The difference was significant in Recall-2 as well ($W=350$, $p<0.05$). All these findings provide enough evidence to support $H2$.

\begin{table}[b]
\renewcommand{\arraystretch}{1.3}
\caption{Descriptive statistics for registration and login time (in seconds)}
\vspace{0.0cm}
\centering
\begin{tabular}{ccccc}
\textbf{Scheme}&Study&Mean&Median&SD\\ 
\hline
\multirow{3}{*}{\textbf{Q-A}}&Registration & $116.6$ & $103$ & $43$ \\
&Recall-1  & $53.9$ & $51$ & $17.8$\\ 
&Recall-2  & $56.9$ & $53$ & $20.5$\\ 
\rule{0pt}{4ex}\multirow{3}{*}{\textbf{Control}}&Registration & $48.1$ & $45.5$ & $27.1$ \\ 
&Recall-1  & $38.1$ & $11.5$ & $70.8$\\ 
&Recall-2  & $43.7$ & $13$ & $55$\\ 
\end{tabular}
\label{tab:time}
\end{table}

\subsection{Hypothesis 3}
\vspace{0.05cm}
\noindent $H3$: \textit{During login in Q-A, the mean time required to enter a letter of the given position would vary significantly for different positions in the answer in both Recall-1 and Recall-2.}
\vspace{0.05cm}

The cognitive effort required to fetch a letter from the third position
of an answer might be higher than entering a letter in the first
position. So, we hypothesized that the mean time required to enter a
letter would vary significantly for different positions in the answer.

We used a Kruskal-Wallis test for Hypothesis 3, which is similar to
ANOVA, but does not make any assumption about the distributions of the
compared samples, and is thus appropriate in this context. The
Kruskal-Wallis test demonstrated that for Recall-1, the mean time
required to enter a letter varied significantly for different positions
in the answers, $\mathcal{X}^{2}(4)= 15.32$, $p<0.01$. However, for
Recall-2, the differences were not statistically significant,
$\mathcal{X}^{2}(4)= 4.79$, $p=0.31$. So, $H3$ is partially supported by
our results.

\subsection{Notable Findings}
\vspace{0.03cm}
\subsubsection{Number of attempts} 
A successful authentication in Q-A requires entering the correct letter
for six questions, but we didn't restrict the number of attempts a user
could make before entering her authentication secret correctly. It
allowed us tracking the number of mistakes a user committed before she
was able to log in successfully. Q-A showed quite promising results in
this respect.

In Recall-1, $82$\% participants succeeded on the first attempt to enter
the correct letter for all six questions. The other participants
made a mistake for only one question and entered the correct letter in
the second attempt. In Recall-2, $73$\% participants succeeded on the
first attempt for all six questions. Of the other six participants
($27$\%) who made a mistake in a question on the first attempt, four
succeeded in the second attempt and the other two required no more than
four attempts before entering the correct letter. One participant made
mistakes in two questions; no other participant made any mistakes in
more than one question in Recall-2.

For the control passwords in Recall-1, $82$\% of participants succeeded
to log in on the first attempt. One participant required $10$ attempts to
log in successfully, as he did not notice that \textit{Caps Lock} was
on. Q-A overcomes this limitation, since the users' responses are not
case sensitive. In Recall-2, $68$\% of participants entered their
control passwords correctly in the first attempt. The participants who
were successfully authenticated after multiple attempts in Recall-2
required four attempts on average with a maximum of eight attempts.

\subsubsection{Registration} 

In Q-A, participants were able to complete registration in less than two minutes on average (see Table~\ref{tab:time}). The mean time to register in the control condition was $48.1$ seconds. A Wilcoxon signed-rank test (appropriate for matched pair of subjects) reveals that there existed a significant difference in registration time between Q-A and control conditions ($W=450.5$, $p<0.01$).

\begin{table}[t]
\renewcommand{\arraystretch}{1.3}
\caption{Questionnaire responses for the usability of Q-A. Scores are
  out of 10. * indicates that the scale was reversed.}  \centering
\begin{tabular}{lcc}
\multicolumn{1}{c}{Questions}&Mode&Median\\ 
\hline
Logging in using Q-A password was easy&$10$&$8.5$\\ 
Q-A passwords are easy to remember&$9$&$9$\\ 
With practice, I could quickly & \multirow{2}{*}{$10$}&\multirow{2}{*}{$9$}\\
enter my Q-A password &&\\
\rule{0pt}{4ex}*I found Q-A too time-consuming (i.e., I &\multirow{2}{*}{$4$}&\multirow{2}{*}{$4.5$}\\
found Q-A to not take too much time) & & \\ 
\rule{0pt}{4ex}*I prefer system-assigned textual   &\multirow{3}{*}{$8$}&\multirow{3}{*}{$7$} \\
passwords to Q-A (i.e., I prefer Q-A to & & \\
system-assigned textual passwords) & &\\ 
\rule{0pt}{4ex}I could easily use Q-A every day&$10$&$8$\\ 
I could easily use Q-A every week&$10$&$9$\\ 
\end{tabular}
\label{tab:usability}
\end{table}

\begin{table}[b]
\renewcommand{\arraystretch}{1.3}
\caption{Questionnaire responses for the applicability of Q-A in
  different online accounts. Scores are out of 10.}
\centering
\begin{tabular}{r@{\hskip 0.8cm}cc}
Online accounts&Mode&Median\\ 
\hline
Bank&$9$&$9$\\ 
Webmail&$9$&$6$\\ 
Social Networking&$5$&$5$\\ 
University Portal&$6$&$6.5$\\ 
E-commerce&$10$&$8$\\ 
\end{tabular}
\label{tab:applicable}
\end{table}

\textbf{Correlations.} We were interested to see if the registration
time for a participant would have any correlation with her login time
and the required number of attempts for a successful authentication in
Q-A. Our results show that the registration time and login time in Q-A
were not strongly correlated in either Recall-1 ($r=0.25$) or Recall-2
($r=0.26$). Also, there existed no strong correlation between the
registration time and the required number of attempts in Q-A for a
successful login in either Recall-1 ($r=-.05$) or in Recall-2 ($r=.20$).

\textbf{Number of mistakes.} During registration in Q-A, answers are
shown as asterisks or dots to reduce the risk of shoulder surfing, and
the users have to re-enter an answer to confirm. Our results show that
users did not make any mistake when confirming the answers. We have
deployed some basic restrictions in our system to guard against poor
answers (e.g., minimum three characters in an answer, no repeat answers
between questions, at least two different letters in an answer). In this
respect, six participants attempted to enter an identical answer for
multiple questions, and when they saw the error message they entered
distinct answers for each question.

\subsection{User Opinion and Perception}

In order to gain an understanding of users' perceptions on the usability and applicability of Q-A, we asked them to answer two sets of Likert scale questions at the end of the second session. We used
ten-point Likert scales, where anchors were included on the bi-polar ends of the scale ($1$ indicating \textit{strong disagreement} and $10$ equalling \textit{strong agreement} with the given statement). We
reversed some of the questions to avoid bias; thus the scores marked with (*) were reversed before calculating the mode and median. A higher score always indicates a more positive result for Q-A. Since Likert scale data are ordinal, it is most appropriate to calculate mode and median for Likert-scale responses~\cite{mean}.

\textbf{Usability.} The perceptions of users on the usability of Q-A are illustrated in Table~\ref{tab:usability}. Users showed a high degree of satisfaction on the usability (e.g., memorability, ease of login) of our scheme, and they preferred Q-A over system-assigned textual passwords. The participants responded positively (i.e., mode and median were higher than neutral) about the ease of using Q-A either weekly or daily. Although they expressed concerns regarding the authentication time in Q-A, they reported that with practice they could log in quickly with this scheme.

\begin{table}[!b]
\renewcommand{\arraystretch}{1.3}
\caption{Questionnaire responses to the requirement of recording
  passwords in real life}
\centering
\begin{tabular}{rccccc}
\multicolumn{1}{c}{\bf Scheme\hspace{0.3cm}}&Never&Rarely&Sometimes&Often&Always\\ 
\hline
{\bf Q-A}\hspace{0.3cm} &$55$\% & $36$\% & $9$\% & $0$\% & $0$\%\\ 
{\bf Control}\hspace{0.3cm} &$0$\% & $14$\% & $46$\% & $36$\% & $4$\%\\ 
\end{tabular}
\label{tab:record}
\end{table}

\textbf{Applicability.} Table~\ref{tab:applicable} shows users' perceptions on the applicability of Q-A in different types of online accounts. Our findings suggest that most of the participants would strongly prefer to use Q-A for online accounts with high security requirements, such as banking and e-commerce accounts. 

The participants were given an open-ended question at the end of second session to express their opinion about Q-A.  In general, the feedbacks were positive and encouraging. They expressed high degree of satisfaction on the security features of Q-A. For example, one participant reported, ``I would use it for banks and other websites, where I use confidential info.'' Several of them were interested to know if Q-A would be deployed commercially.

\textbf{Password recording.}  All the participants reported that they did not write down their control nor Q-A passwords for this study. We also asked them if they would require writing down their Q-A or control passwords if they would use them in real life. The results (see Table~\ref{tab:record}) are quite promising in this respect: $55$\% of participants reported that they would \textit{never} need to record their Q-A password, while $36$\% of participants would \textit{rarely} need to. Only two ($9$\%) of the participants mentioned that they would \textit{sometimes} require to write down their Q-A password.

For control passwords, $46$\% of participants would \textit{sometimes} require and $36$\% of participants would \textit{often} need to record their password.

\section{Discussion}\label{disc}
The login success rate for Q-A was found to be $100$\% in both Recall-1
and Recall-2. Here we discuss the anticipated factors that might have
played an important role to gain such high a performance during
authentication.

\begin{itemize}
\item Q-A asks for already known information that are related to the
  long term memory of a user. So, the user does not have to memorize any
  artificial information as her authentication secret.
\item Q-A provides users with the flexibility to choose any six
  questions from a set of twenty, which they find most applicable to them.
\item The basic restrictions (see~\S\ref{sec_adv}) should have motivated
  users to enter the correct answer instead of a weak placeholder one
  (`Aab'). Real answers to cognitive questions have been shown to be
  more memorable than placeholder answers~\cite{no_sec}.
\item Being specific with the question helps to accurately recall the
  answer in Q-A. For example: instead of asking ``Where did your father
  and mother meet?'', we would ask ``In what city or town did your
  father and mother meet?'
\item Q-A makes users focus on fetching a letter from the given position
  in her answer, which reduces the chance of typing mistakes that occur
  when they have to enter the whole answer.
\item Q-A is case-insensitive, while confusion between capital and
  lowercase letters has been shown to be a prominent reason for making
  mistakes when answering cognitive questions~\cite{qa04}.
\end{itemize}

\subsection{Limitations and Ecological Validity} 
Our participants were young and university educated, which represents a large number of frequent Web users, but may not generalize to the entire population. As the study was performed in a lab setting, we were only able to gather data from $22$ participants. 

Although field studies may provide superior ecological validity, lab studies have been preferred to examine brain-powered memorability of passwords~\cite{ecology13}. Moreover, lab studies have the advantage of taking place in a controlled setting, which helps to establish performance bounds and figure out whether field tests are worthwhile in future research. 
 
Our results for login time are conservative since they reflect initial use. As noted in existing research~\cite{design_space}, login time likely decreases with frequent use of a scheme. Chiasson et al.~\cite{passpoint3} find that users get more attentive while entering passwords in a lab study, which can contribute to less login error and higher login time than the casual login attempts in real-life scenario. Further long-term studies in a real life setting could provide additional insights on the training effect.

\section{Conclusion and Future Work}\label{conc}

Our study on Q-A finds that cognitive questions possess good possibilities in primary authentication in terms of both usability and security. We note that in addition to providing sufficient entropy to prevent online guessing attacks, Q-A is resilient to both internal and external observation attacks. Memorability in Q-A was found to be $100$\% in our study. Users' feedbacks reveal that they are highly satisfied with the memorability and security of this scheme, which would be most appropriate for online accounts with high security requirements (e.g., online banking, e-commerce~\cite{pw_diary}). 

Now that we have observed promising memorability for single Q-A password in a lab setting, it seems reasonable to plan for a field study with larger and more diverse populations in order to examine if cognitive questions can provide a usable solution to the memorability and interference problems with multiple passwords.




\bibliographystyle{IEEEtran}      
\bibliography{IEEEabrv,refs}



\end{document}